\documentclass[
 reprint,
 amsmath,amssymb,
 aps,
 pra,
]{revtex4-1}

\usepackage{graphicx}
\usepackage{dcolumn}
\usepackage{bm}
\usepackage{siunitx}
\usepackage{hyperref}
\usepackage{subfig}
\usepackage{svg}
\usepackage{caption}
\newcommand{\R}{\mathbb{R}}
\renewcommand{\Im}{\text{Im}}
\renewcommand{\Re}{\text{Re}}
\newcommand{\adj}{\text{adj}}
\newcommand{\Graph}{\mathcal{G}}
\newcommand{\Circuit}{\mathcal{C}}
\newcommand{\Vertices}{\mathcal{V}}
\newcommand{\LRC}{LRC}

\usepackage{xspace}
\newcommand{\tlinecirc}{1a\xspace}
\newcommand{\purcellcirc}{1b\xspace}
\newcommand{\tworescirc}{1c\xspace}

\begin{document}

\title{Computational modeling of decay and hybridization in superconducting circuits}

\author{Michael~G.~Scheer}\email{mgscheer@gmail.com}
\thanks{These authors contributed equally.}
\author{Maxwell~B.~Block}\email{mblock@berkeley.edu}
\thanks{These authors contributed equally.}

\affiliation{Rigetti Computing, 2919 Seventh St, Berkeley CA 94710}

\date{\today}

\begin{abstract}
We present a framework for modeling superconducting circuits that integrates classical microwave analysis with circuit quantization. Our framework enables the calculation of the lossy eigenmodes of superconducting circuits, and we demonstrate the method by analyzing several circuits relevant to multiplexed, Purcell filtered transmon readout architectures. We show that the transmon relaxation times obtained by our method agree with the established approximation $T_1 \approx C/\Re[Y(i\omega_q)]$ away from environmental resonances and do not vanish on resonance. We also show that the hybridization of the modes in the readout circuit is highly sensitive to the bandwidth of the Purcell filter.
\end{abstract}

\maketitle

\section{Introduction}\label{sec:intro}
\begin{figure}[h]
    \centering
    \includegraphics[trim={6.8cm 9cm 6.8cm 4cm},clip]{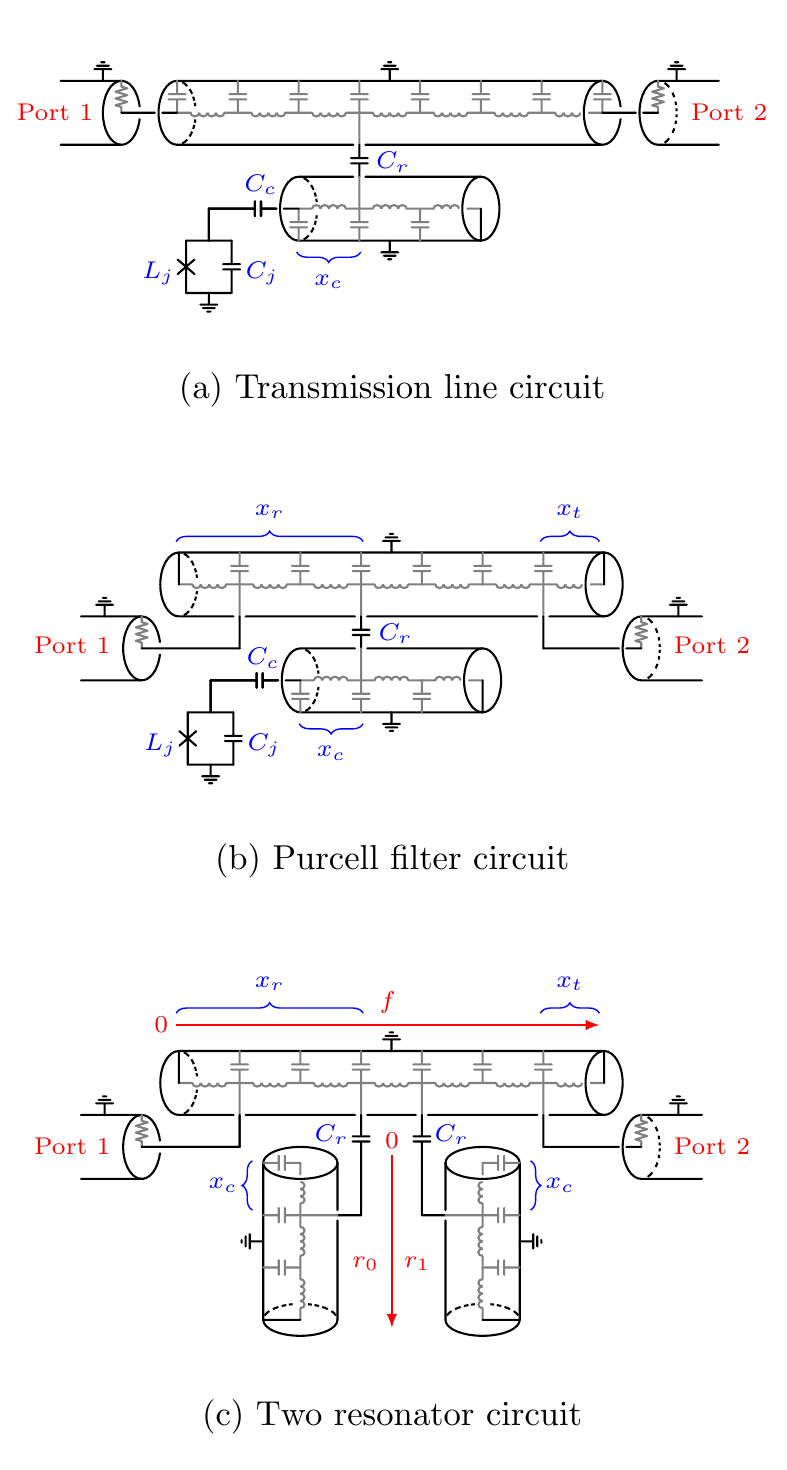}
    \caption{The three circuits analyzed in section \ref{sec:modeling}. The ideal continuous circuits are in black and possible discrete approximations are in grey. In (c), $f$, $r_0$, and $r_1$ indicate coordinate systems along the Purcell filter and the two resonators. Capacitances and lengths are labelled in blue while ports and coordinate systems are labelled in red.}
    \label{fig:circs}
\end{figure}

Many proposals for superconducting qubit devices capable of achieving quantum advantage require innovations such as high qubit connectivity \cite{Song2017, Versluis2017, Reagor2018, Hong2019}, multiplexed and filtered readout \cite{Reed2010, Jeffrey2014, Kelly2015, Walter2017, Heinsoo2018}, and 3D integration \cite{OBrien2017, Vahidpour2017, Mutus2017, Bejanin2016}. As a consequence, the linear electromagnetic aspects of superconducting devices are becoming increasingly complex and important to model accurately. Many superconducting qubit experiments have used physically motivated circuit models consisting of lumped elements and transmission lines to model their devices \cite{Schuster2007, Houck2008, Reed2010, Jeffrey2014}. Typically, the parameters for these models are derived from electrostatic or magnetostatic simulations. An alternative approach is to infer a circuit model from a transfer function computed with a full-wave electromagnetic solver \cite{Nigg2012, Solgun2014, Solgun2015, Solgun2017}. Methods for computing the Hamiltonian for a circuit model and approximating the relaxation time of a transmon due to its environment have also been developed \cite{Esteve1986, Devoret1996, Martinis2003, Burkard2004, Burkard2005, DiVincenzo2006, Vool2017}.

While there is significant literature on creating and analyzing circuit models, a scalable and rigorous approach to understanding the lossy eigenmodes of a complex superconducting device has yet to emerge. In this paper, we present a simple framework that fills this gap. We use a particular class of state-space models we refer to as Positive Second Order (PSO) models to represent the linear aspects of superconducting circuits. These models have been studied extensively because of their ability to represent the dynamics of arbitrary $RLC$ circuits \cite{Sheehan2003, Bai2005}. We explain how to construct PSO models for superconducting circuits and demonstrate that they can be used to understand the Purcell effect \cite{Purcell1946} and mode hybridization.

We begin in section \ref{sec:PSO} by defining PSO models and reviewing how to determine their eigenmodes, complex frequencies, and transfer functions. In section \ref{sec:operations} we describe a method for combining PSO models. In sections \ref{sec:circuits} and \ref{sec:tline-boundaries} we provide a derivation of the PSO model for the node-flux coordinates \cite{Devoret1996} of a lumped circuit, possibly including semi-infinite transmission lines. Finally, in section \ref{sec:lagrangian} we note that in the absence of loss a PSO model is equivalent to a Lagrangian written in node-flux coordinates. Consequently, a quantum model including Josephson Junction degrees of freedom can be easily obtained from a PSO model.

To demonstrate the utility of this framework, we apply the results of section \ref{sec:theory} to investigate two important problems in superconducting circuit design: the Purcell effect \cite{Purcell1946} and readout crosstalk. For the Purcell effect, we analyze in section \ref{sec:purcell} the circuits shown in Figs. \tlinecirc and \purcellcirc, which model a dispersive readout system for a transmon qubit without and with a Purcell filter, respectively \cite{Reed2010, Jeffrey2014, Kelly2015, Walter2017, Heinsoo2018}. We observe the expected increase in qubit relaxation time due to the Purcell filter when the qubit is detuned from the modes of the environment. We also compare our results with the well established approximation for qubit relaxation time \cite{Esteve1986, Martinis2003, Houck2008, Neeley2008}
\begin{equation}\label{eqn:c_re_y}
T_1 \approx \frac{C}{\Re[Y(i\omega_q)]}
\end{equation} where $Y$ is the admittance of the environment seen by the qubit at the qubit's transition frequency and $C$ is the effective total capacitance of the qubit. When the qubit is detuned from the modes of the environment we see very good agreement with Eq. (\ref{eqn:c_re_y}). If the qubit is near a resonance, Eq. (\ref{eqn:c_re_y}) predicts a vanishing qubit relaxation time, but the PSO model predicts a finite qubit relaxation time, in agreement with \cite{Malekakhlagh2016}.

To investigate readout crosstalk, we analyze the circuit shown in Fig. \tworescirc which models part of a multiplexed dispersive readout system for transmon qubits with a Purcell filter. Since a major component of readout crosstalk comes from mode hybridization, we consider the spatial eigenmode distributions as a function of the bandwidth of the Purcell filter. When the Purcell filter is completely lossless, the modes are highly coupled, and they decouple as the Purcell filter becomes lossy. Interestingly, we observe that the frequencies and decay rates vary non-monotonically with the bandwidth of the Purcell filter.

\section{Theory}\label{sec:theory}
\subsection{Positive Second Order Models}\label{sec:PSO}
A Positive Second Order (PSO) model is a continuous mapping from input vectors to output vectors of the form
\begin{align}
K\Phi + G\dot{\Phi} + C\ddot{\Phi} &= PD\label{eqn:agc-model-input}\\
V &= P^{T}\dot{\Phi}\label{eqn:agc-model-output}
\end{align}
where $D$ and $V$ are the input and output and $\Phi$ are some internal degrees of freedom. The matrices $K$, $G$, and $C$ are real and positive semidefinite and $P$ is an arbitrary real matrix. $K$ and $C$ give rise to conservative dynamics, while $G$ is responsible for decay. $P$ represents the way in which the inputs and outputs couple to the internal degrees of freedom. We will often represent a PSO model by the tuple $(K, G, C, P)$.

Eqs. (\ref{eqn:agc-model-input}, \ref{eqn:agc-model-output}) are equivalent to the state-space model
\begin{align}
\begin{pmatrix}
G & K\\-I & 0
\end{pmatrix}
\Psi
+
\begin{pmatrix}
C & 0\\0 & I
\end{pmatrix}
\dot{\Psi}
&=
\begin{pmatrix}
P\\0
\end{pmatrix}
D\label{eqn:agc-state-space}\\
V &= \begin{pmatrix}
P^T & 0
\end{pmatrix}
\Psi
\end{align}
where $\Psi = [\dot{\Phi}, \Phi]^T$. The eigenmodes of a PSO model are the solutions of Eq. (\ref{eqn:agc-model-input}) with $D=0$. From Eq. (\ref{eqn:agc-state-space}) it is clear that these can be found by solving the generalized eigenvalue problem
\begin{equation}\label{eqn:generalized_eigenvalue}
\begin{pmatrix}
-G & -K\\I & 0
\end{pmatrix}
\begin{pmatrix}
v_{\dot{\Phi}}\\v_{\Phi}
\end{pmatrix} = \lambda
\begin{pmatrix}
C & 0\\0 & I
\end{pmatrix}
\begin{pmatrix}
v_{\dot{\Phi}}\\v_{\Phi}
\end{pmatrix}
\end{equation}
for the eigenmode $v_{\Phi}$ with complex frequency $\lambda$. The frequency and decay rate of the eigenmode are $\Im(\lambda)/2\pi$ and $-2\Re(\lambda)/2\pi$ respectively.

If we take the Laplace transform of Eqs. (\ref{eqn:agc-model-input}, \ref{eqn:agc-model-output}) using $s$ as the complex Laplace variable, we find $V(s) = Z(s) D(s)$ where
\begin{equation}\label{eqn:impedance}
Z(s) = P^T\left(\frac{K}{s} + G + Cs\right)^{-1}P
\end{equation}
is the transfer function of the model.

\subsection{Operations on PSO models}\label{sec:operations}
We will now discuss three useful operations on PSO models. First, we consider a coordinate transformation. Let $M = (K, G, C, P)$ be a PSO model. If we left-multiply Eq. (\ref{eqn:agc-model-input}) by an invertible matrix $U$ and define $\Phi' = (U^{T})^{-1}\Phi$, Eqs. (\ref{eqn:agc-model-input}, \ref{eqn:agc-model-output}) become
\begin{align}
UKU^{T}\Phi' + UGU^{T}\dot{\Phi}' + UCU^{T}\ddot{\Phi}' &= UPD\\
V &= (UP)^T\Phi'.
\end{align}
We therefore obtain a new PSO model
\begin{equation}
M' = (UKU^T, UGU^T, UCU^T, UP).
\end{equation}
Since $U$ is invertible, $M$ and $M'$ represent the same dynamics and have the same complex frequencies and transfer function. Their eigenmodes are related by 
\begin{equation}
v_{\Phi'} = (U^T)^{-1}v_{\Phi}
\end{equation}
where $v_{\Phi}$ and $v_{\Phi'}$ are corresponding eigenmodes of $M$ and $M'$ respectively.

Next, we consider the union of PSO models. Suppose we have two PSO models $M_0 = (K_0, G_0, C_0, P_0)$ and $M_1 = (K_1, G_1, C_1, P_1)$ with inputs, outputs and coordinates $(D_0, V_0, \Phi_0)$ and $(D_1, V_1, \Phi_1)$. We form a new PSO model $M_2 = (K_2, G_2, C_2, P_2)$ with
\begin{equation}
X_2 = \begin{pmatrix}
X_0 & 0\\
0 & X_1
\end{pmatrix}
\end{equation}
for $X = K, G, C, P$, with input $D_2 = [D_0, D_1]^T$, output $V_2 = [V_0, V_1]^T$ and coordinates $\Phi_2 = [\Phi_0, \Phi_1]^T$. $M_2$ represents the same dynamics as $M_0$ and $M_1$ did separately, but in one model.

Finally, we consider constraint satisfaction. Suppose we have a PSO model $M = (K, G, C, P)$ and in addition to Eqs. (\ref{eqn:agc-model-input}, \ref{eqn:agc-model-output}) we require
\begin{equation}\label{eqn:constraint}
Y^T\Phi = 0
\end{equation}
for some real matrix $Y$. In general, this is not consistent with Eqs. (\ref{eqn:agc-model-input}, \ref{eqn:agc-model-output}) if the inputs $D$ are arbitrary functions of time. However if $D$ is chosen in such a way that Eq. (\ref{eqn:constraint}) is satisfied, we can find a reduced PSO model consistent with Eq. (\ref{eqn:constraint}).

Let $Z$ be a matrix whose columns form a basis for the null space of $Y^T$. Then Eq. (\ref{eqn:constraint}) is equivalent to
\begin{equation}\label{eqn:substitute-constraint}
\Phi = Z \Phi'
\end{equation} for some vector $\Phi'$. Without loss of generality, suppose $Y$ is full rank so that $(Y, Z)$ is invertible. If we substitute Eq. (\ref{eqn:substitute-constraint}) into Eqs. (\ref{eqn:agc-model-input}, \ref{eqn:agc-model-output}) and left-multiply Eq. (\ref{eqn:agc-model-input}) by $(Y, Z)^T$ we find
\begin{align}
Y^T KZ\Phi' + Y^T GZ\dot{\Phi}' + Y^T CZ\ddot{\Phi}' &= Y^T PD\label{eqn:agc-constraint-for-input}\\
Z^T KZ\Phi' + Z^T GZ\dot{\Phi}' + Z^T CZ\ddot{\Phi}' &= Z^T PD\label{eqn:agc-constraint-input}\\
V &= P^{T}Z\dot{\Phi}'\label{eqn:agc-constraint-output}.
\end{align}
Eqs. (\ref{eqn:agc-constraint-input}, \ref{eqn:agc-constraint-output}) represent a new PSO model
\begin{equation}
M' = (Z^T K Z, Z^T G Z, Z^T C Z, Z^T P)
\end{equation} while Eq. (\ref{eqn:agc-constraint-for-input}) shows the constraint that must be satisfied by $D$ so that Eqs. (\ref{eqn:agc-model-input}, \ref{eqn:agc-model-output}, \ref{eqn:constraint}) can be simultaneously satisfied.

By using union and constraint satisfaction, several PSO models can be combined to create a new PSO model with different dynamics. This is a convenient way to build complex PSO models from simple building blocks. For example, in section \ref{sec:modeling} we use this approach to build complex circuit models from transmission lines and lumped elements.

\subsection{Circuits}\label{sec:circuits}
We will now show that the equations of motion for a circuit consisting of inductors, capacitors, and resistors can be represented by a PSO model. An example circuit and corresponding PSO model are given in appendix \ref{app:example-circuit}. We formalize a circuit $\Circuit$ as a complete graph $\Graph$ with vertices $\Vertices$ along with three symmetric functions from $\Vertices\times \Vertices \to \R^+$ denoted $k$, $g$, $c$. The vertices represent locations where charge can accumulate and the edges represent pathways for currents to flow in the form of a parallel inductor, resistor, and capacitor. The three functions $k$, $g$, and $c$ assign to each edge its lumped element circuit values in the form of inverse inductance, conductance, and capacitance. We use inverse inductance and conductance rather than inductance and resistance so that a value of $0$ is an open circuit for all three functions \cite{Devgan2000, Sheehan2003}. The circuit can be driven by a time-dependent current on each edge $d_I(v,v')$ as well as a time-dependent magnetic flux through any loop $d_\Phi(v, \dots, v')$. Note that both $d_I$ and $d_\Phi$ switch sign when their arguments are reversed.

Following \cite{Devoret1996, Vool2017}, we use the node flux coordinates $\phi$ to represent the state of the circuit. Specifically, we define $\phi(v, v')$ to be the time integral of the voltage difference between $v$ and $v'$. In these coordinates, the current conservation equation at vertex $v$ is
\begin{equation}\label{eqn:abstract_equations_of_motion}
\sum_{v'\in \Vertices} (k\phi + g\dot{\phi}+ c\ddot{\phi} - d_I)(v,v') = 0.
\end{equation}

This gives us a set of $|\Vertices|$ equations of motion with $|\Vertices|^2$ free variables $\phi(v,v')$. However, neither the variables nor the equations are independent. Since $\phi$ is defined as the time integral of a voltage difference, $\phi$ is anti-symmetric. Furthermore, Faraday's law implies that $\phi$ sums around any loop to the enclosed magnetic flux $d_\Phi$. As a result of these constraints, there are only $|\Vertices|-1$ independent variables. Additionally, since both $\phi$ and $d_I$ are anti-symmetric while $k$, $g$, and $c$ are symmetric, the left hand sides of Eqs. (\ref{eqn:abstract_equations_of_motion}) sum identically to $0$. This means that there are only $|\Vertices|-1$ independent equations. We now proceed to derive equations of motion with independent equations and independent variables.

Consider a directed rooted spanning tree $S$ for $\Graph$ with root $r$ and with all edges oriented towards the root \cite{Devoret1996,Vool2017}. Let the edges of $S$ be $e_{1}, \dots, e_{|\Vertices|-1}$ and let $\Phi$ be a length $|\Vertices|-1$ vector with $\Phi_i = \phi(e_i)$. Note that these coordinates are independent because $S$ contains no loops. For any two vertices $v$ and $v'$ let the unique path through $S$ from $v$ to $v'$ be called $\rho(v,v')$. Define a function $l: \Vertices \to \{-1,0,1\}^{|\Vertices|-1}$ so that for all $v$
\begin{equation}\label{eqn:l(v)}
l(v)\cdot \Phi = \sum_{e\in \rho(v,r)} \phi(e).
\end{equation}
For any two vertices $v$ and $v'$ the concatenation of the paths $\rho(r,v)$, $(v,v')$ and $\rho(v',r)$ forms a loop. For brevity, we will write $d_\Phi(v,v')$ for the magnetic flux in this loop. Note that the function $d_\Phi$ is determined by its values on these loops. We have
\begin{equation}
-l(v)\cdot \Phi + \phi(v,v') + l(v')\cdot \Phi = d_\Phi(v,v')
\end{equation}
and therefore
\begin{equation}
\phi(v, v') = (l(v) - l(v'))\cdot \Phi + d_\Phi(v,v').
\end{equation}
Define $m(v, v') = l(v) - l(v')$ so that
\begin{equation}\label{eqn:express_phi}
\phi(v,v') = m(v,v')\cdot \Phi + d_\Phi(v,v').
\end{equation}
Next, choose an ordering for the vertices $v_1, \dots, v_n \in \Vertices$ and define a $|\Vertices|\times (|\Vertices|-1)$ matrix $\hat{K}$ with
\begin{equation}
\hat{K}_{ij} = \sum_{v'\in \Vertices} k(v_i, v')m(v_i, v')_j
\end{equation}
and similarly define $\hat{G}$ and $\hat{C}$.
Additionally, let $\hat{D}$ be a length $|\Vertices|$ time-dependent vector with
\begin{equation}\label{eqn:d_hat}
\hat{D}_i = -\sum_{v' \in \Vertices} (kd_\Phi + g\dot{d_\Phi} + c\ddot{d_\Phi} - d_I)(v_i, v').
\end{equation}
Substituting (\ref{eqn:express_phi}) into (\ref{eqn:abstract_equations_of_motion}) yields
\begin{equation}\label{eqn:hat_matrix_equations_of_motion}
\hat{K}\Phi + \hat{G}\dot{\Phi} + \hat{C}\ddot{\Phi} = \hat{D}.
\end{equation}

We have now solved the problem of redundant coordinates, but we still have one redundant equation. In principle, one could remove any single equation. However, since we aim to produce a PSO model, care must be taken to create positive semidefinite $K$, $G$, and $C$ matrices. Let $T$ be a $(|\Vertices|-1) \times |\Vertices|$ matrix with columns $l(v_i)$. We left-multiply Eq. (\ref{eqn:hat_matrix_equations_of_motion}) by $T$ and note that no information is lost since $T$ is full-rank. We now have
\begin{equation}\label{eqn:matrix_equations_of_motion}
K\Phi + G\dot{\Phi} + C\ddot{\Phi} = D
\end{equation}
where $K = T\hat{K}$ and similarly for $G$, $C$ and $D$. Note
\begin{align}
K_{\alpha\beta} &= \sum_{i} T_{\alpha i}\hat{K}_{i \beta}\\
&= \sum_{i,j} l(v_i)_{\alpha} k(v_i, v_j) m(v_i, v_j)_{\beta}\\
&= \frac{1}{2}\sum_{i,j} m(v_i, v_j)_{\alpha}k(v_i, v_j) m(v_i, v_j)_{\beta}\label{line:symmetry}\\
&= \frac{1}{2}\sum_{e\in \Vertices\times\Vertices} m(e)_\alpha k(e) m(e)_{\beta}\label{line:explicit_circuit_matrix}
\end{align}
where line (\ref{line:symmetry}) follows by the symmetry of $k$. Furthermore for any vector $v$,
\begin{equation}
v^T K v = \frac{1}{2}\sum_{e\in \Vertices\times\Vertices} k(e) (m(e) \cdot v)^2
\geq 0.
\end{equation}
Since these equations also hold for $G$ and $C$, we see that all three matrices are positive semidefinite. As a result, equation (\ref{eqn:matrix_equations_of_motion}) corresponds to a PSO model $(K, G, C, I)$ where the input and output are $D$ and $\dot{\Phi}$.

Now, we consider the case that the only drives present are current drives on some list $L$ of edges we will refer to as \textit{ports} \cite{Pozar2011}. Let $L=(p_1, \dots, p_n)$ and let $\Phi^L$ and $D^L$ be length $n$ vectors with $\Phi^L_i = \phi(p_i)$ and $D^L_i = d_I(p_i)$. Let $P$ be a $(|\Vertices|-1) \times n$ matrix with columns $m(p_i)$. Equation (\ref{eqn:express_phi}) implies that $\Phi^L = P^{T}\Phi$. Furthermore, Eq. (\ref{eqn:d_hat}) with only current drives can be written as a sum over ports
\begin{equation}
\hat{D}_i = \sum_{(v_j, v_k)\in L} d_I(v_j, v_k)(\delta_{ij} - \delta_{ik})
\end{equation}
so
\begin{equation}
D = T\hat{D} = PD^L.
\end{equation}
If we define $V^L = \dot{\Phi}^L$, we have
\begin{align}
K\Phi + G\dot{\Phi} + C\ddot{\Phi} &= PD^{L}\label{eqn:matrix_equations_with_ports_input}\\
V^L &= P^{T}\dot{\Phi}.\label{eqn:matrix_equations_with_ports_output}
\end{align}
These equations correspond to a PSO model $(K, G, C, P)$ with input and output $D^L$ and $V^L$. Note that in this context the transfer function $Z$ in Eq. (\ref{eqn:impedance}) is the impedance matrix.

\subsection{Transmission line boundary conditions}\label{sec:tline-boundaries}
We have shown that lumped circuits can be represented by PSO models. We now demonstrate that lumped circuits with semi-infinite transmission lines also admit a PSO model representation. Suppose that a semi-infinite transmission line of characteristic impedance $Z_0$ is terminated in a port $p\in L$ of the circuit $\Circuit$ so that the drive current $d_I(p)$ is the transmission line current at the port. Let $z \geq 0$ indicate position along the transmission line with the port at $z=0$, and let $V(z)$ and $I(z)$ indicate the voltage and current on the transmission line. The telegrapher equations imply
\begin{align}
I(z) = \frac{2}{Z_0}V^{+}(z) - \frac{1}{Z_0}V(z)
\end{align}
where $V^{+}$ is the traveling voltage wave moving towards the port \cite{Pozar2011}. Since the port is at $z=0$, we have $\dot{\phi}(p) = V(0)$ and so
\begin{equation}
d_I(p) = I(0) = \frac{2}{Z_0}V^{+}(0) - \frac{1}{Z_0}\dot{\phi}(p)
\end{equation}
Using this relation, we see that
\begin{equation}
(k\phi + g\dot{\phi} + c\ddot{\phi}-d_I)(p) = (k\phi + g'\dot{\phi} + c\ddot{\phi}-d_I')(p)
\end{equation}
where
\begin{align}
g'(p) &= g(p) + \frac{1}{Z_0}\\
d_I'(p) &= \frac{2}{Z_0}V^{+}(0).
\end{align}
Therefore, to include a transmission line in a circuit model, it suffices to add a resistor $Z_0$ and a current drive $\frac{2}{Z_0}V^{+}(0)$ to the edge to which it is connected \cite{Yurke1984}. Dissipation of energy in this resistor represents energy radiated into the transmission line.

\subsection{Lagrangian formulation}\label{sec:lagrangian}
If $G = 0$, Eq. (\ref{eqn:agc-model-input}) is the Euler-Lagrange equation for the Lagrangian \cite{Burkard2004, Burkard2005}
\begin{equation}
\mathcal{L}(\Phi, \dot{\Phi}) = \frac{1}{2}\dot{\Phi}^T C \dot{\Phi} - \frac{1}{2}\Phi^T K \Phi + \Phi^T PD^{L}.
\end{equation}
Note that the symmetry of $K$ and $C$ is necessary for this statement to be true. If the PSO model represents the equations of motion for a circuit as described in section \ref{sec:circuits} then the addition of a Josephson Junction with Josephson energy $E_J$ to an edge $e$ can be achieved by adding
\begin{equation}
E_J \cos(m(e)\cdot \Phi + d_\Phi(e))    
\end{equation}
to the Lagrangian.

\section{Modeling superconducting circuits}\label{sec:modeling}
We now apply the results of section \ref{sec:theory} to analyze the three circuits depicted in Fig. \ref{fig:circs}. The first system consists of a transmon qubit coupled to a quarter-wave transmission line resonator, itself coupled to a semi-infinite transmission line. The second system is similar to the first, but additionally has a half-wave bandpass Purcell filter \cite{Reed2010, Jeffrey2014, Kelly2015, Walter2017, Heinsoo2018}. The last system consists of two quarter-wave resonators coupled to a common Purcell filter. All couplings are mediated by lumped capacitors, and losses through semi-infinite transmission lines are modeled with resistors as described in section \ref{sec:tline-boundaries}. The parameters of the circuits shown in Fig. \ref{fig:circs} are summarized in appendix \ref{app:parameters}.

In order to apply the results in section \ref{sec:circuits}, we must approximate the finite transmission lines and the transmon qubits with discrete circuits. The transmission lines are approximated with $LC$ ladder circuits \cite{Pozar2011} and we show evidence in appendix \ref{app:convergence} that the error due to this approximation is small. Because of their small anharmonicity, we approximate the transmons as parallel $LC$ resonators. A rigorous justification of this approximation can be found in \cite{Malekakhlagh2017}. Given the parallel $LC$ resonator, series capacitor, and $LC$ ladder as building blocks, one can construct PSO models for the circuits shown in Fig. \ref{fig:circs} using the operations described in section \ref{sec:operations}.

\subsection{Radiative loss}\label{sec:purcell}
\begin{figure}
    \centering
    \includegraphics[scale=0.4]{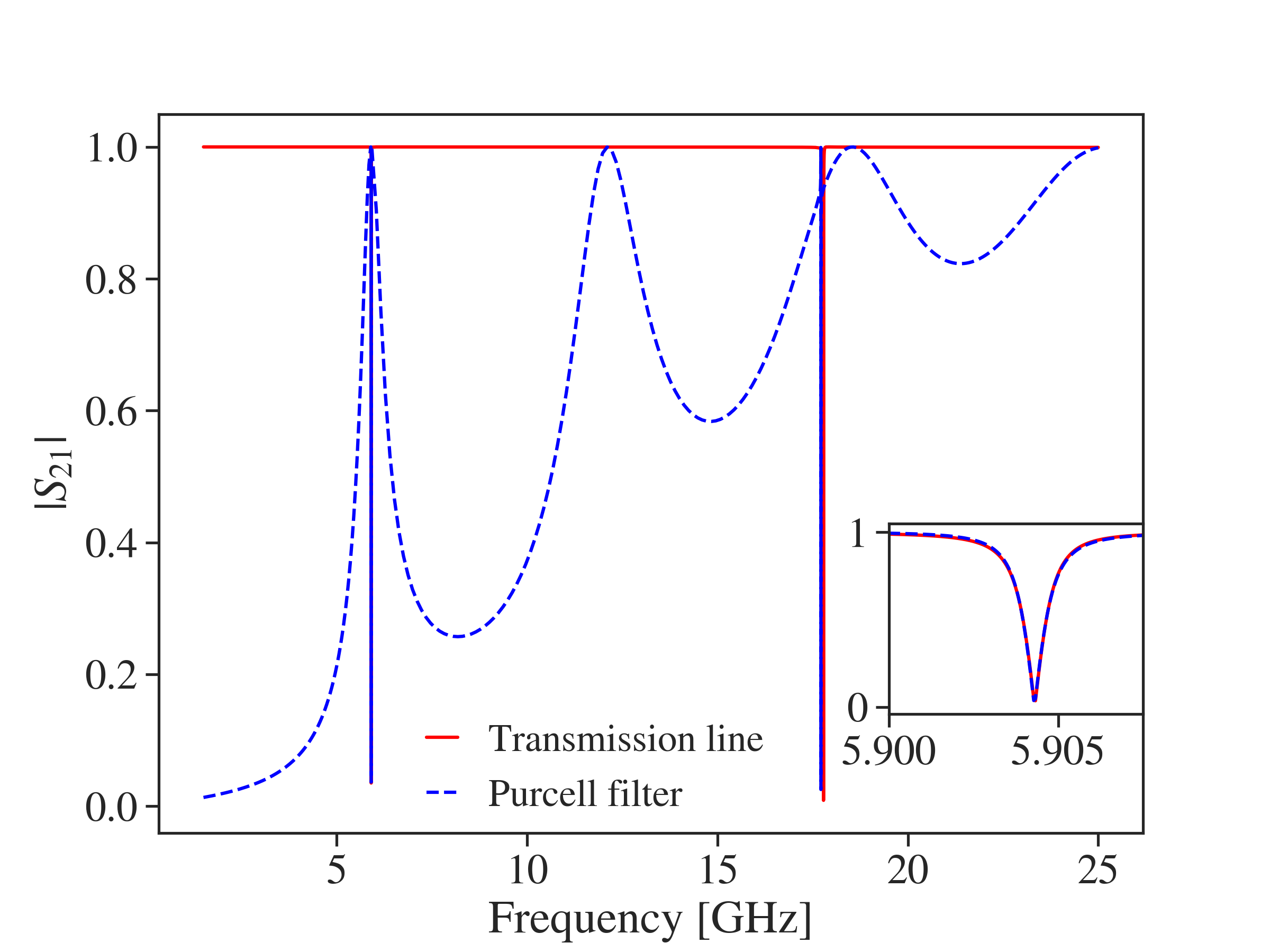}
    \caption{Magnitude of the transmission coefficient for the circuits depicted in Figs. \tlinecirc and \purcellcirc. The inset shows the transmission near the lowest resonance of each system. The parameters of the models were chosen so that the lowest resonances have nearly identical frequency and linewidth.}
    \label{fig:mag_s12}
\end{figure}

\begin{figure}
    \centering
    \includegraphics[scale=0.4]{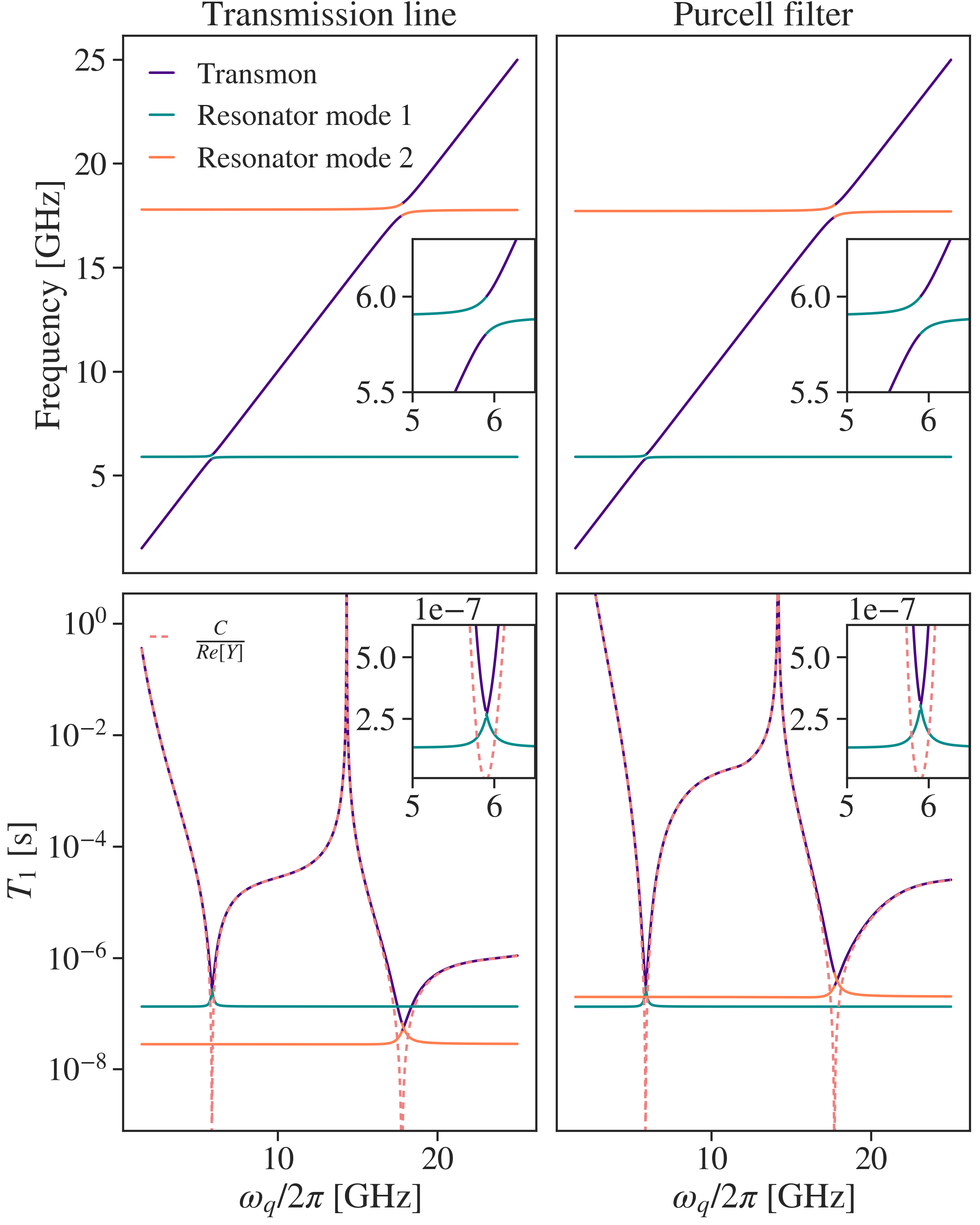}
    \caption{Frequency (top) and $T_1$ (bottom) for the transmon mode and the lowest two resonator modes as a function of $\omega_q$ for the circuits depicted in Figs. \tlinecirc (left) and \purcellcirc (right). All curves are computed from PSO model complex frequencies, except the dashed curve which shows the single port admittance model for transmon relaxation time. The top insets show the avoided crossing between the transmon mode and the lowest resonator mode and the bottom insets show the corresponding relaxation times. The single port admittance model matches the complex frequencies except near resonance where it vashishes. In contrast, the PSO model predicts that the transmon and resonator relaxation times match on resonance.}
    \label{fig:complex_frequencies}
\end{figure}

In this section, we compare the the complex frequencies of the circuits shown in Figs. \tlinecirc and \purcellcirc for a range of transmon frequencies. Fig. \ref{fig:mag_s12} shows the magnitude of the transmission coefficient between ports $1$ and $2$ for both circuits with the transmon detuned from the resonator modes. The scattering parameters $S(s)$ are computed using Eq. (\ref{eqn:impedance}) along with the relation \cite{Pozar2011}
\begin{equation}
S(s) = (Z(s) + Z_0I)^{-1}(Z(s)-Z_0I)
\end{equation}
where $Z_0$ is characteristic impedance of the transmission lines. As demonstrated in Fig. \ref{fig:mag_s12}, the parameters of the two circuits were chosen so that their resonators have nearly identical frequency and linewidth.

We solve Eq. (\ref{eqn:generalized_eigenvalue}) to compute the complex frequencies of each system as a function of the inductance of the transmon. Fig. \ref{fig:complex_frequencies} shows the frequencies and relaxation times for the transmon and the two lowest resonator modes for both circuits. The lower panels also include the single port admittance approximation \cite{Esteve1986, Martinis2003, Houck2008, Neeley2008} for the relaxation time of the transmon mode, which we now briefly review.

Let the transmon consist of a parallel capacitance $C_j$ and inductance $L_j$. Let $Y_e$ be the admittance of the circuit as seen by the transmon. We compute this admittance exactly using analytical expressions for lumped circuit elements and transmission lines. As shown in appendix
\ref{app:admittance_matrix}, $Y_e$ has leading capacitive and inductive terms with coefficients $C_e$ and $\frac{1}{L_e}$, respectively. The transmon is then modeled as a parallel $\LRC$ resonator where $\frac{1}{L} = \frac{1}{L_j} + \frac{1}{L_e}$, $C = C_j + C_e$, and $R=\frac{1}{\Re{Y_e(i\omega_q)}}$, where $\omega_q = \frac{1}{\sqrt{L C}}$ is the bare angular frequency of the transmon. The transmon relaxation time is approximated as the $RC$ constant for this resonator,
\begin{equation}\label{eqn:admittance_t1}
T_1(\omega_q) \approx \frac{C}{\Re[Y_e(i\omega_q)]}.
\end{equation}

Note that Eq. (\ref{eqn:admittance_t1}) is sensitive to the values of the environmental reactances $C_e$ and $\frac{1}{L_e}$ both through $C$ and $\omega_q$, so it is important to estimate them accurately. In appendix \ref{app:ce_le}, we provide numerical evidence that $C_e \approx C_c$ and $\frac{1}{L_e} \approx 0$ for the circuits depicted in Figs. \tlinecirc and \purcellcirc.

We find that the two relaxation time calculations are in excellent agreement away from resonance. Near resonance with the first resonator mode, where the top panels show avoided crossings, the single port admittance model predicts vanishing relaxation times. In contrast, the PSO model predicts that the transmon relaxation time matches that of the resonator mode. This is in agreement with \cite{Malekakhlagh2016} in which it was shown that the decay rates of the transmon and resonator are approximately equal when the coupling is much larger than the resonator decay rate, as is the case here.

Comparing the bottom two panels of Fig. \ref{fig:complex_frequencies}, we see that when the transmon is detuned from environmental modes, its relaxation time is roughly two orders of magnitude higher in the Purcell filter circuit than in the transmission line circuit. This demonstrates that the Purcell filter is effective in increasing the relaxation time of the transmon while the resonator frequency and linewidth are held fixed. Additionally, both circuits demonstrate significantly enhanced transmon relaxation times when $\omega_q/2\pi \approx \nu/2 x_c$ where $\nu$ is the transmission line propagation speed and $x_c$ is the distance between the quarter-wave resonator short and its coupler to the environment. At this frequency, the short is transformed to appear at the coupler, so the transmon mode has a node there and does not radiate.

\subsection{Hybridization}\label{sec:hybridization}
\begin{figure}
    \centering
    \includegraphics[scale=0.4]{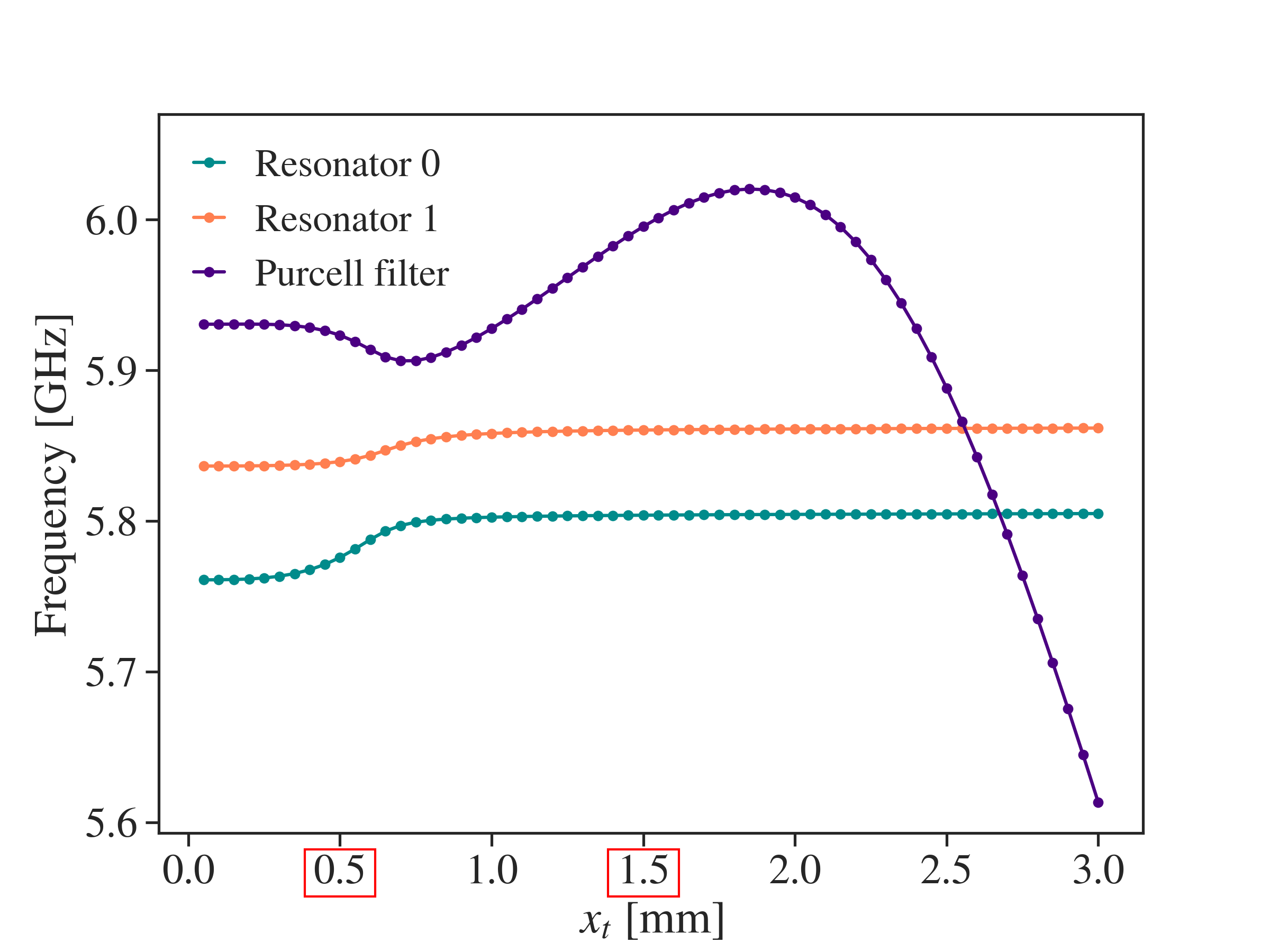}
    \caption{Frequencies of the modes of the circuit depicted in Fig. \tworescirc as a function of $x_t$. For $x_t$ from $\SI{0.5}{\milli\meter}$ to $\SI{1.0}{\milli\meter}$ the frequencies shift closer as loss induced decoupling reduces the mode repulsion. The red boxes indicate the values of $x_t$ used in Figs. \ref{fig:pfilters_s12} and \ref{fig:hybridization}.}
    \label{fig:frequencies_v_stubs}
\end{figure}

\begin{figure}
    \centering
    \includegraphics[scale=0.4]{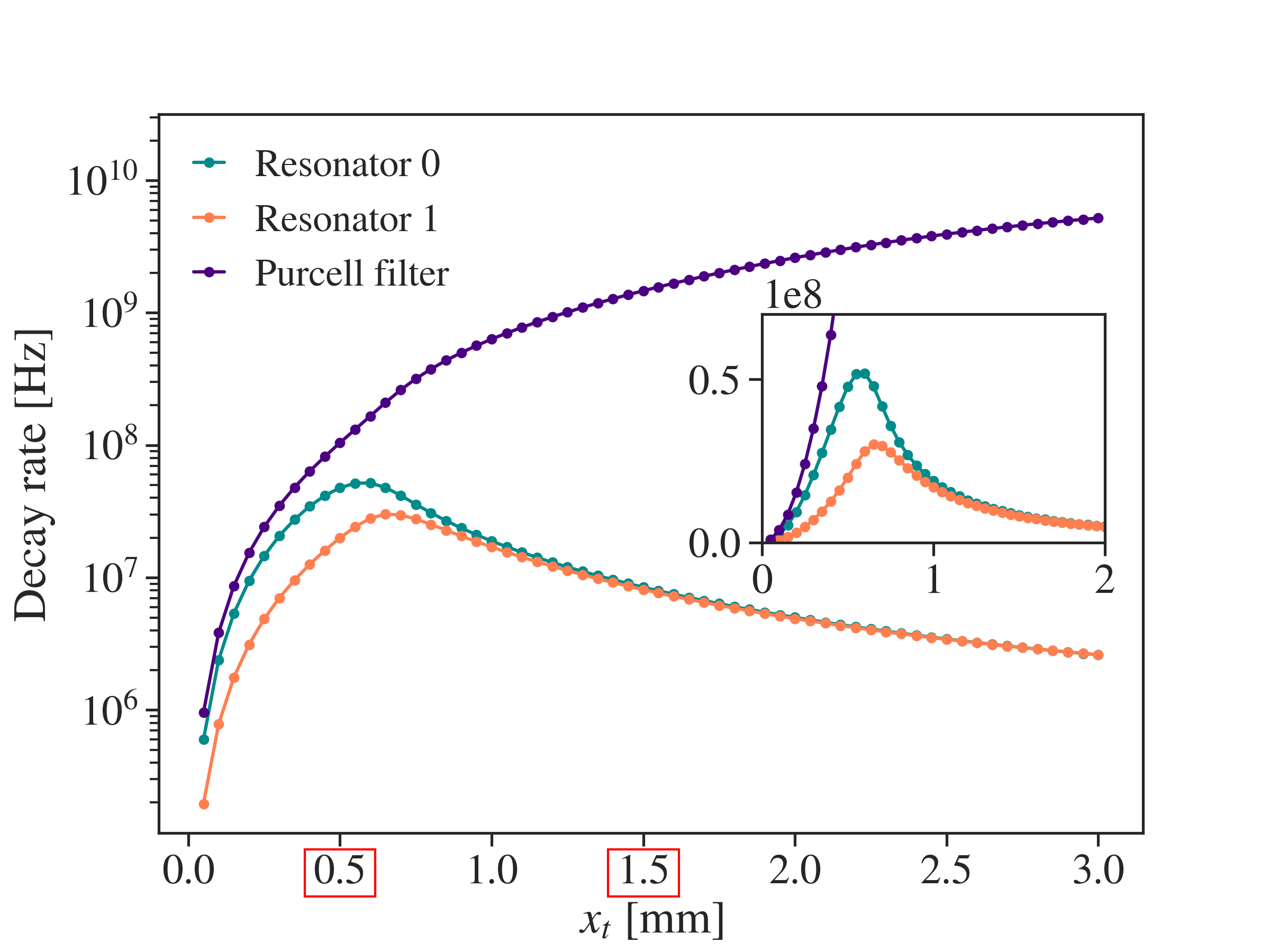}
    \caption{Decay rates of the modes of the circuit depicted in Fig. \tworescirc as a function of $x_t$. The inset shows a subset of the data with a linear $y$ axis. The Purcell filter decay rate increases monotonically with $x_t$. The resonator decay rates attain maxima and then decline as the resonators decouple from the Purcell filter. The red boxes indicate the values of $x_t$ used in Figs. \ref{fig:pfilters_s12} and \ref{fig:hybridization}.}
    \label{fig:loss_rates_vs_stubs}
\end{figure}

\begin{figure}
    \centering
    \includegraphics[scale=0.4]{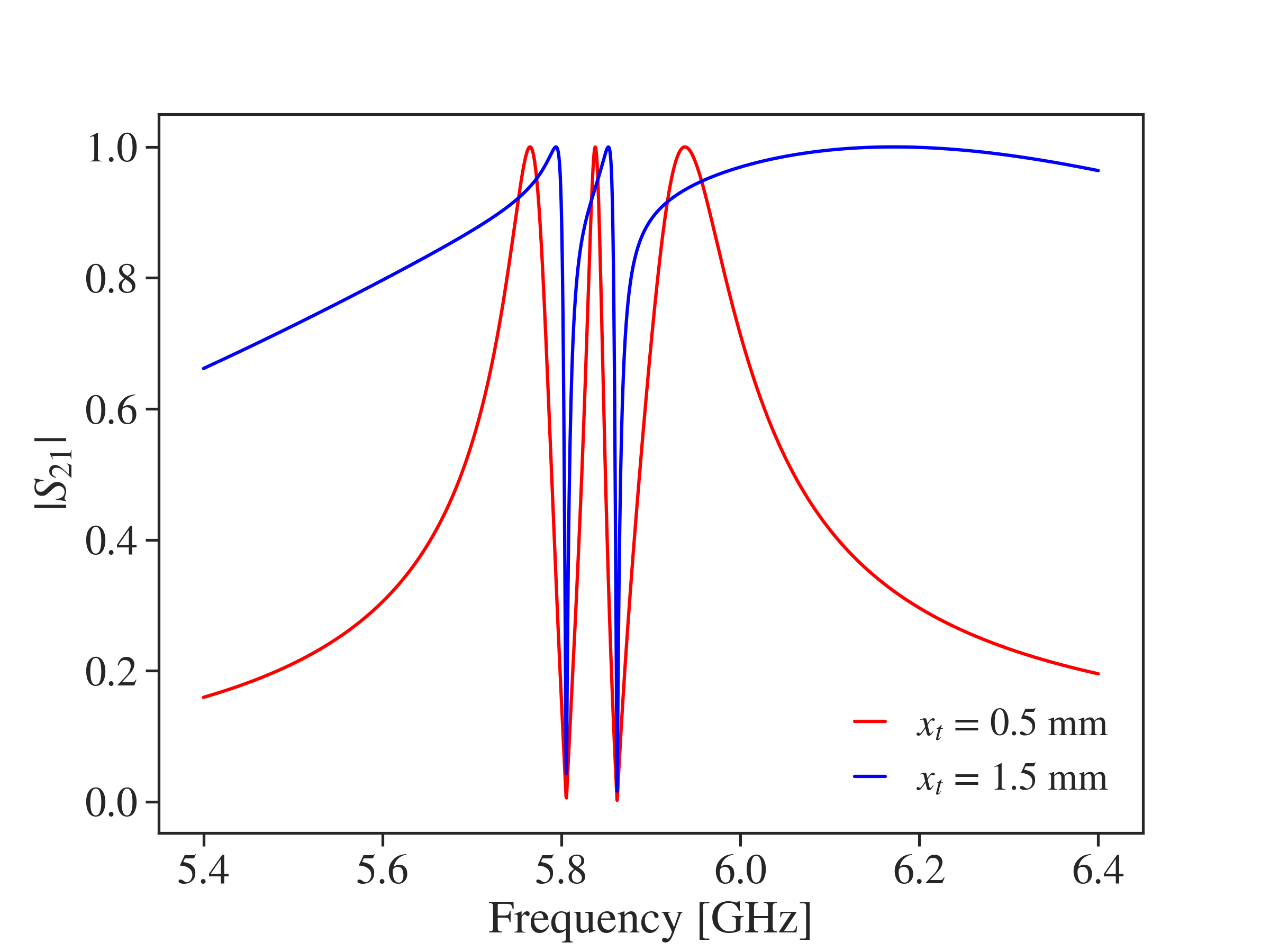}
    \caption{Magnitude of the transmission coefficient for the circuit depicted in Fig. \tworescirc for $x_t=\SI{0.5}{\milli\meter}$ and $\SI{1.5}{\milli\meter}.$ For $x_t=\SI{0.5}{\milli\meter}$, the Purcell filter is narrow-band and the resonators are lossy. For $x_t=\SI{1.5}{\milli\meter}$ the Purcell filter is broad-band, the resonators have narrower linewidth, and all modes have slightly higher frequencies.}
    \label{fig:pfilters_s12}
\end{figure}

\begin{figure}
    \centering
    \includegraphics[scale=0.4]{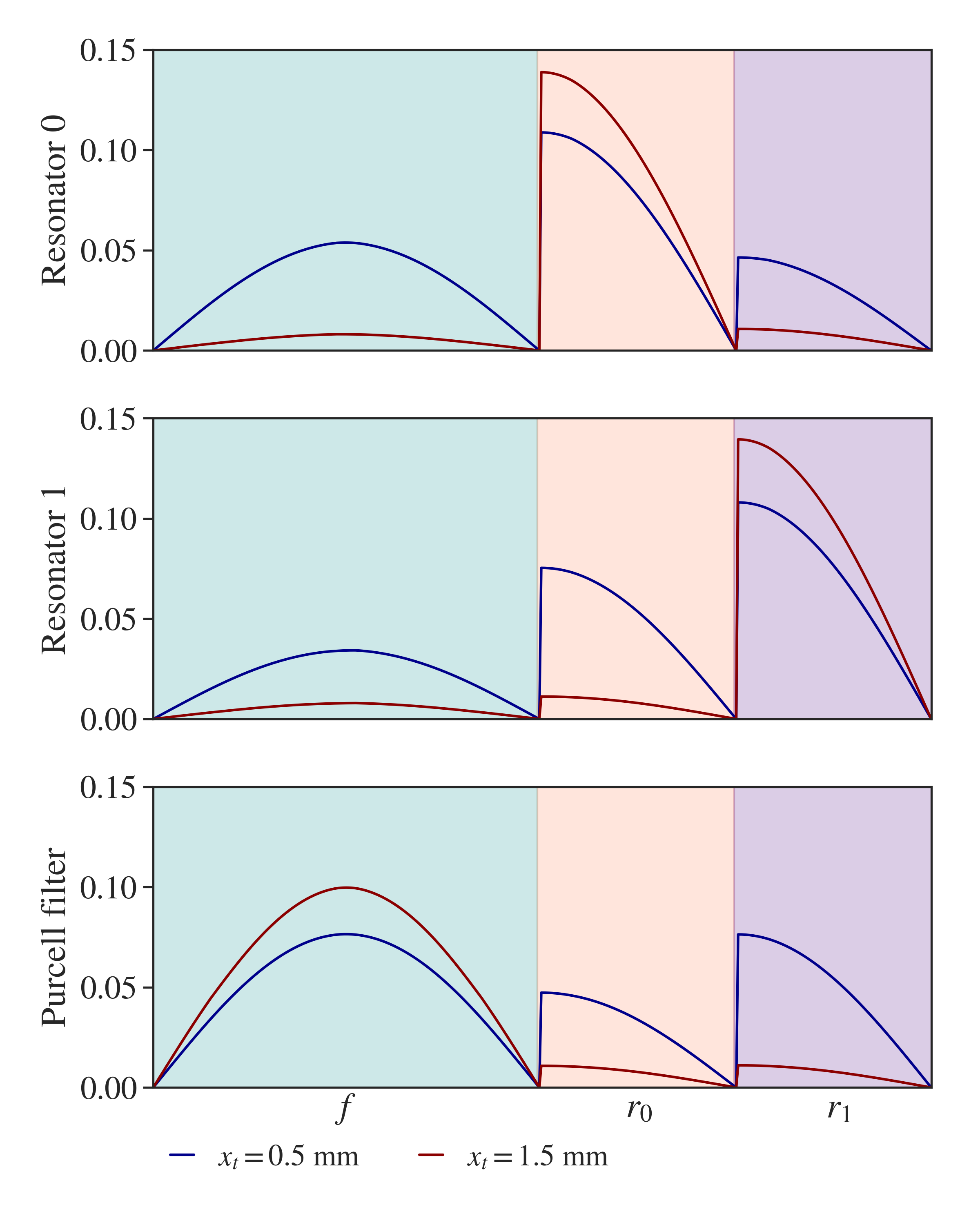}
    \caption{Magnitude of the normalized eigenmodes of the circuit depicted in Fig. \tworescirc for $x_t = \SI{0.5}{\milli\meter}$ and $\SI{1.5}{\milli\meter}$. The subspaces $f$, $r_0$, and $r_1$ indicated in Fig. \tworescirc are color-coded. The eigenmodes for $x_t = \SI{0.5}{\milli\meter}$ are significantly more hybridized than for $x_t = \SI{1.5}{\milli\meter}$.}
    \label{fig:hybridization}
\end{figure}

\begin{figure}
    \centering
    \includegraphics[scale=0.4]{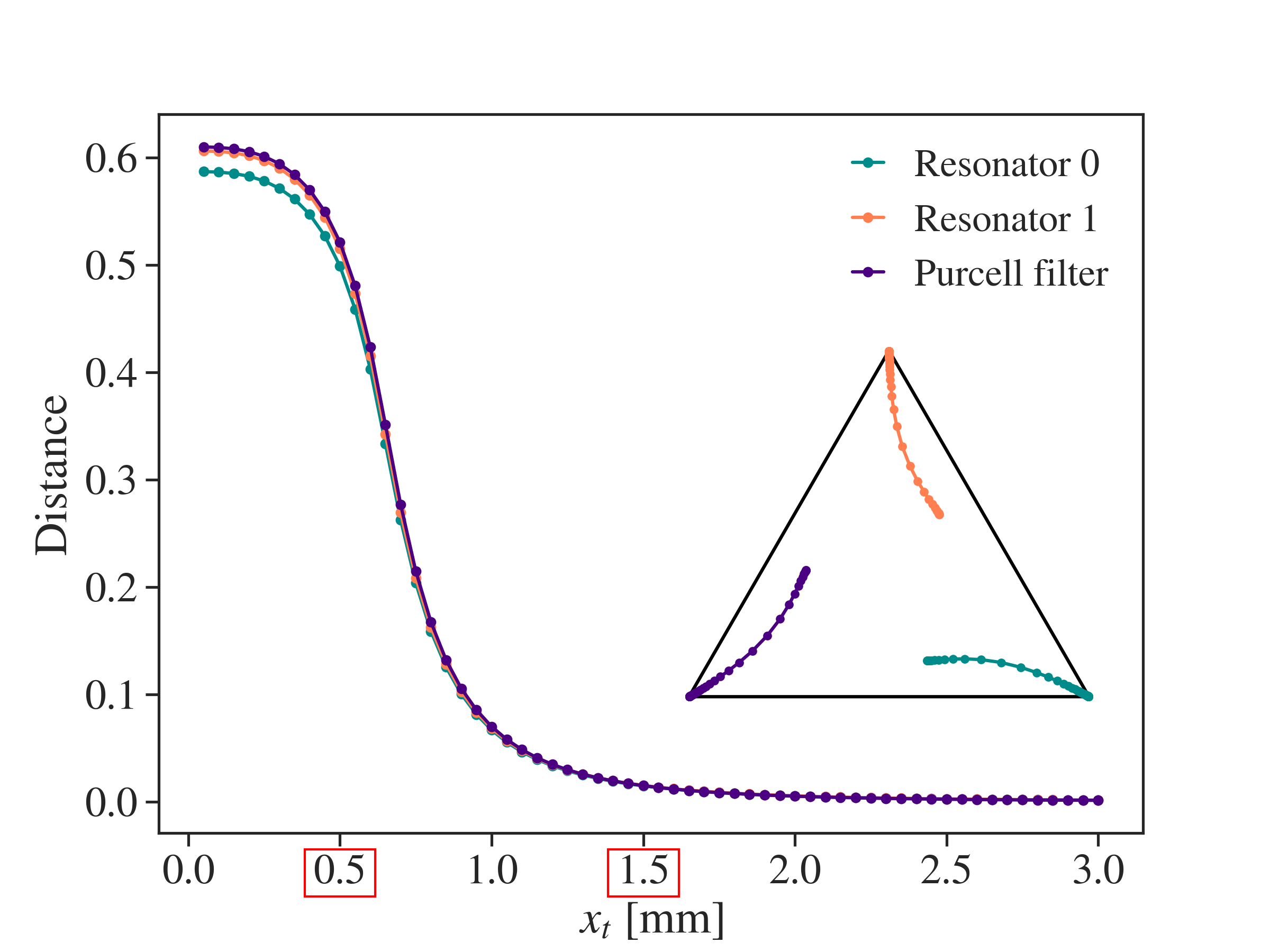}
    \caption{Distance between each regional support vector and the corresponding standard basis vector for the circuit depicted in Fig. \tworescirc. The distance goes to $0$ for large $x_t$ as a result of loss induced decoupling. The inset shows the regional support vectors on the plane determined by the standard basis. It is interesting to note the asymmetrical structure of the hybridization in this system which can also be observed in Fig. \ref{fig:hybridization}.}
    \label{fig:barycentric_overlaps}
\end{figure}

\newcommand{\xtmin}{\SI{50}{\micro\meter}}
\newcommand{\xtmax}{\SI{3.0}{\milli\meter}}
\newcommand{\xtsmall}{\SI{0.5}{\milli\meter}}
\newcommand{\xtlarge}{\SI{1.5}{\milli\meter}}
In this section, we investigate the hybridization of the modes of the circuit depicted in Fig. \tworescirc as a function of the distance $x_t$ from the Purcell filter shorts to the transmission lines. We compute the complex frequencies and eigenmodes of the system for $x_t$ ranging from $\xtmin$ to $\xtmax$ and show the frequencies and decay rates in Figs. \ref{fig:frequencies_v_stubs} and \ref{fig:loss_rates_vs_stubs}. The Purcell filter decay rate increases with $x_t$ as the transmission line couplings move from electric field nodes towards the electric field anti-node. This trend can also be observed in the bandwidth of the Purcell filter scattering parameters, shown in Fig. \ref{fig:pfilters_s12}.

Several interesting effects can be observed in Figs. \ref{fig:frequencies_v_stubs} and \ref{fig:loss_rates_vs_stubs}. First, the resonator decay rates have a distinctive peak around $\xtsmall$. Second, resonator 0 has a significantly higher maximum decay rate than resonator 1 despite being more detuned from the Purcell filter at the relevant $x_t$ and being otherwise identical. Third, the resonator frequencies rise and the Purcell filter frequency drops near $\xtsmall$. Finally, the Purcell filter's frequency rises again and then asymptotically drops for large $x_t$.

The key to understanding these effects is to examine the spatial structure of the eigenmodes. Figure \ref{fig:hybridization} shows the magnitudes of the normalized eigenmodes for $x_t=\xtsmall$ and $\xtlarge$ by their projections onto the $f$, $r_0$, and $r_1$ subspaces indicated in Fig. \tworescirc. The resonator eigenmodes for $x_t = \xtsmall$ have significant support on the Purcell filter, allowing them to decay into the environment. In contrast, the resonator eigenmodes for $x_t = \xtlarge$ are largely isolated and therefore protected from decay. These conclusions are in agreement with Fig. \ref{fig:loss_rates_vs_stubs}.

In order to visualize the eigenmode structure at more than a handfull of $x_t$ values, we employ a low dimensional representation of the eigenmodes. For each normalized eigenmode, we compute the norm-squared of its projection onto the $f$, $r_0$, and $r_1$ subspaces and consider these as vectors in $\R^3$. We refer to these vectors as \textit{regional support vectors}. Since each has non-negative components summing to $1$, they lie in the convex hull of the standard basis. In the absence of coupling, they would in fact be the standard basis since each mode would only be supported on a single region.

Figure \ref{fig:barycentric_overlaps} shows the distance between each regional support vector and its corresponding standard basis vector as a function of $x_t$. The eigenmode hybridization monotonically decreases with $x_t$ and therefore Purcell filter decay rate, an effect we call \textit{loss induced decoupling}. This effect can be understood by analogy to three parallel $\LRC$ resonators capacitively coupled in series. If the resistance of the central resonator is reduced to $0$ its decay rate goes to $\infty$ and it becomes a short to ground. In this limit, all three modes are decoupled. The competing trends of increasing Purcell filter decay rate and loss induced decoupling explain the rise and fall of the resonator decay rates observed in Fig. \ref{fig:loss_rates_vs_stubs}. The attraction of resonator frequencies to the Purcell filter frequency for small $x_t$ can also be explained by loss induced decoupling. As $x_t$ increases and the modes decouple, the Purcell filter ceases to repel the resonators.

The inset of Fig. \ref{fig:barycentric_overlaps} shows the regional support vectors projected onto the plane passing through the standard basis. We see that the hybridization in the system is asymmetrical, which is counter to the naive expectation that both resonator 0 and resonator 1 would predominantly couple to the Purcell filter. This explains why resonator 0 has a higher maximum decay rate. Although it is more detuned, it actually has higher support on the Purcell filter than does resonator 1.

The decline in Purcell filter frequency for large $x_t$ is analogous to the drop in frequency of a parallel $\LRC$ resonator as $R\to 0$. It is interesting to note that for large $x_t$ the Purcell filter frequency passes through both resonators without visible avoided crossings, confirming that the coupling there is negligible. Although it is difficult to confirm, we speculate that the rise of the Purcell filter frequency for moderate $x_t$ is due to the partial confinement of the mode between the impedance mismatches where the transmission lines attach.

\section{Conclusion}
We have presented a framework for modeling the linear aspects of superconducting circuits that enables the computation of the lossy eigenmodes of a complex device. The framework is based on the analysis of PSO models, a class of state-space models that describe $RLC$ circuit dynamics. We have shown that PSO models can be used to compute transmon relaxation times and to explain experimentally relevant variations in resonator frequencies and linewidths.

\section{Acknowledgements}
We are grateful to Blake Johnson, Michael Selvanayagam, Andre Melo, Hakan T\"ureci, Eyob Sete, Nikolas Tezak, and Daniel Girshovich for useful conversations. This work was funded by Rigetti \& Co Inc., dba Rigetti Computing.

\appendix

\section{Example circuit}\label{app:example-circuit}
\begin{figure}
     \centering
     \includegraphics[trim={6.5cm 12cm 6.6cm 12.5cm},clip]{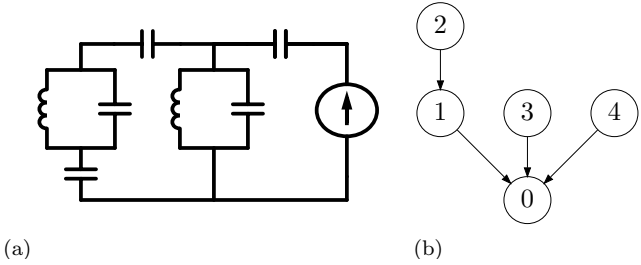}
    \caption{(a) An example circuit consisting of capacitors, inductors, and a current drive. (b) A possible spanning tree for the circuit. The spanning tree establishes the node fluxes that will be used as coordinates in the PSO model.}
     \label{fig:example_circuit}
\end{figure}

In this appendix we provide an example of a circuit and its corresponding PSO model as described in section \ref{sec:circuits}. Fig. \ref{fig:example_circuit} shows a circuit along with a particular choice of spanning tree. The vertices are $\Vertices = \{0, 1, 2, 3, 4\}$ and Eq. (\ref{eqn:l(v)}) implies that $l$ is given by

\begin{align}
l(0) &= [0, 0, 0, 0]\\
l(1) &= [1, 0, 0, 0]\\
l(2) &= [1, 1, 0, 0]\\
l(3) &= [0, 0, 1, 0]\\
l(4) &= [0, 0, 0, 1].
\end{align}

Using Eq. (\ref{line:explicit_circuit_matrix}), we find

\begin{widetext}
\begin{align}
K &= \begin{pmatrix}
0 & 0 & 0 & 0\\
0 & k(2,1) & 0 & 0\\
0 & 0 & k(3,2) & 0\\
0 & 0 & 0 & 0\\
\end{pmatrix}\\
G &= \begin{pmatrix}
0 & 0 & 0 & 0\\
0 & 0 & 0 & 0\\
0 & 0 & 0 & 0\\
0 & 0 & 0 & 0\\
\end{pmatrix}\\
C &= \begin{pmatrix}
c(1,0) + c(2,3) & c(2,3) & -c(2,3) & 0\\
c(2,3) & c(2,1) + c(2,3) & -c(2,3) & 0\\
-c(2,3) & -c(2,3) & c(3,0) + c(2,3) + c(3,4) & -c(3,4)\\
0 & 0 & -c(3,4) & c(3,4)\\
\end{pmatrix}\\
P &= \begin{pmatrix}
0 \\ 0 \\ 0 \\ 1
\end{pmatrix}.
\end{align}
\end{widetext}

\section{Circuit Parameters}\label{app:parameters}
Table \ref{tab:params} shows the numerical values of all circuit components. The referents of capacitances, inductances, and lengths are indicated in Fig. \ref{fig:circs}. In the circuits depicted in Figs. \tlinecirc and \purcellcirc, $l_r$ is the length of the resonator. In the circuit depicted in Fig. \tworescirc, $l_0$ and $l_1$ are the lengths of the left and right resonators respectively. In the circuits depicted in Figs. \purcellcirc and \tworescirc, $l_f$ is the length of the Purcell filter.
\begin{table}
\begin{tabular}{ |c|c|c|c| } 
    \hline
    Parameter & Circ a. & Circ b. & Circ c. \\
    \hline
    $l_r$ & \SI{4.99171}{\milli\meter} & \SI{5.03299}{\milli\meter} & - \\ 
    $l_0$ & - & - & \SI{5.1}{\milli\meter} \\
    $l_1$ & - & - & \SI{5.05}{\milli\meter} \\
    $l_f$ & - & \SI{10.2522}{\milli\meter} & \SI{10.1}{\milli\meter} \\
    $x_c$ & \SI{800}{\micro\meter} & \SI{800}{\micro\meter} & \SI{800}{\micro\meter} \\
    $x_t$ & - & \SI{880}{\micro \meter} & various \\
    $x_r$ & - & \SI{5}{\milli \meter} & \SI{4.8}{\milli \meter} \\
    $C_r$ & \SI{10}{\femto\farad} & \SI{2.6}{\femto\farad} & \SI{12}{\femto\farad} \\
    $C_c$ & \SI{7}{\femto\farad} & \SI{7}{\femto\farad} & - \\
    $C_j$ & \SI{100}{\femto\farad} & \SI{100}{\femto\farad} & - \\
    $L_j$ & \SI{10}{\nano\henry} & \SI{10}{\nano\henry} & - \\
    $\nu$ & \SI{1.2e8}{\meter\per\second} & \SI{1.2e8}{\meter\per\second} & \SI{1.2e8}{\meter\per\second} \\
    $Z_0$ & \SI{50}{\ohm} & \SI{50}{\ohm} & \SI{50}{\ohm} \\
    \hline
\end{tabular}
\caption{Parameters for the circuits in Fig. \ref{fig:circs}}
\label{tab:params}
\end{table}

\section{Admittance matrix}\label{app:admittance_matrix}
In this appendix, we derive an expression for the admittance matrix of a PSO model that is used to explain the single port admittance model in section \ref{sec:purcell}. Suppose that $P$ is full rank so it is possible to make a coordinate transformation such that
\begin{equation}\label{eqn:normal_ports}
P = \begin{pmatrix}
I \\ 0
\end{pmatrix}
\end{equation}
where $I$ is the $n \times n$ identity matrix. Let
\begin{equation}
\frac{K}{s} + G + Cs =
\begin{pmatrix}
Y_1 & Y_2^T\\
Y_2 & Y_3
\end{pmatrix}
\end{equation}
where $Y_1$ is an $n\times n$ matrix. From Eqs. (\ref{eqn:impedance}) and (\ref{eqn:normal_ports}) we have
\begin{align}
Z &= P^T
\begin{pmatrix}
Y_1 & Y_2^T\\
Y_2 & Y_3
\end{pmatrix}
^{-1}P \\
&= (Y_1 - Y_2^TY_3^{-1}Y_2)^{-1}
\end{align}
and so
\begin{equation}
Y = Y_1 - Y_2^TY_3^{-1}Y_2
\end{equation}
is the admittance matrix of the PSO model.
Recalling that the inverse of a matrix is its adjugate divided by its determinant, we have
\begin{equation}\label{eqn:adjugate_determinant}
Y_{ij} = (Y_1)_{ij} -\sum_{a, b}
\frac{\adj(Y_3)_{ab}}{\det(Y_3)} (Y_2)_{ai}(Y_2)_{bj}.
\end{equation}
Each term in this sum is a rational function of $s$ with numerator degree one greater than denominator degree, and where the denominator is divisible by $s$. Using partial fraction decomposition, each rational function can be written in the form
\begin{equation}
\frac{\tilde{K}^{ab}_{ij}}{s} + \tilde{G}^{ab}_{ij} + \tilde{C}^{ab}_{ij}s + q^{ab}_{ij}(s) 
\end{equation}
where $q^{ab}_{ij}$ is a rational function with numerator degree less than denominator degree. This gives
\begin{equation}\label{eqn:admittance}
Y_{ij}(s) = \frac{\tilde{K}_{ij}}{s} + \tilde{G}_{ij} + \tilde{C}_{ij}s + \sum_{a, b} q^{ab}_{ij}(s)
\end{equation}
where $\tilde{K}_{ij} = (P^T K P)_{ij} + \sum_{a, b} \tilde{K}^{ab}_{ij}$ and similarly for $\tilde{G}$ and $\tilde{C}$. The resonant behavior is encoded in the $q^{ab}_{ij}$ functions, while the asymptotic behavior is encoded in the $\tilde{K}$, $\tilde{G}$, and $\tilde{C}$ matrices.

\section{Environmental reactance}\label{app:ce_le}
\begin{figure}
    \centering
    \includegraphics[scale=0.4]{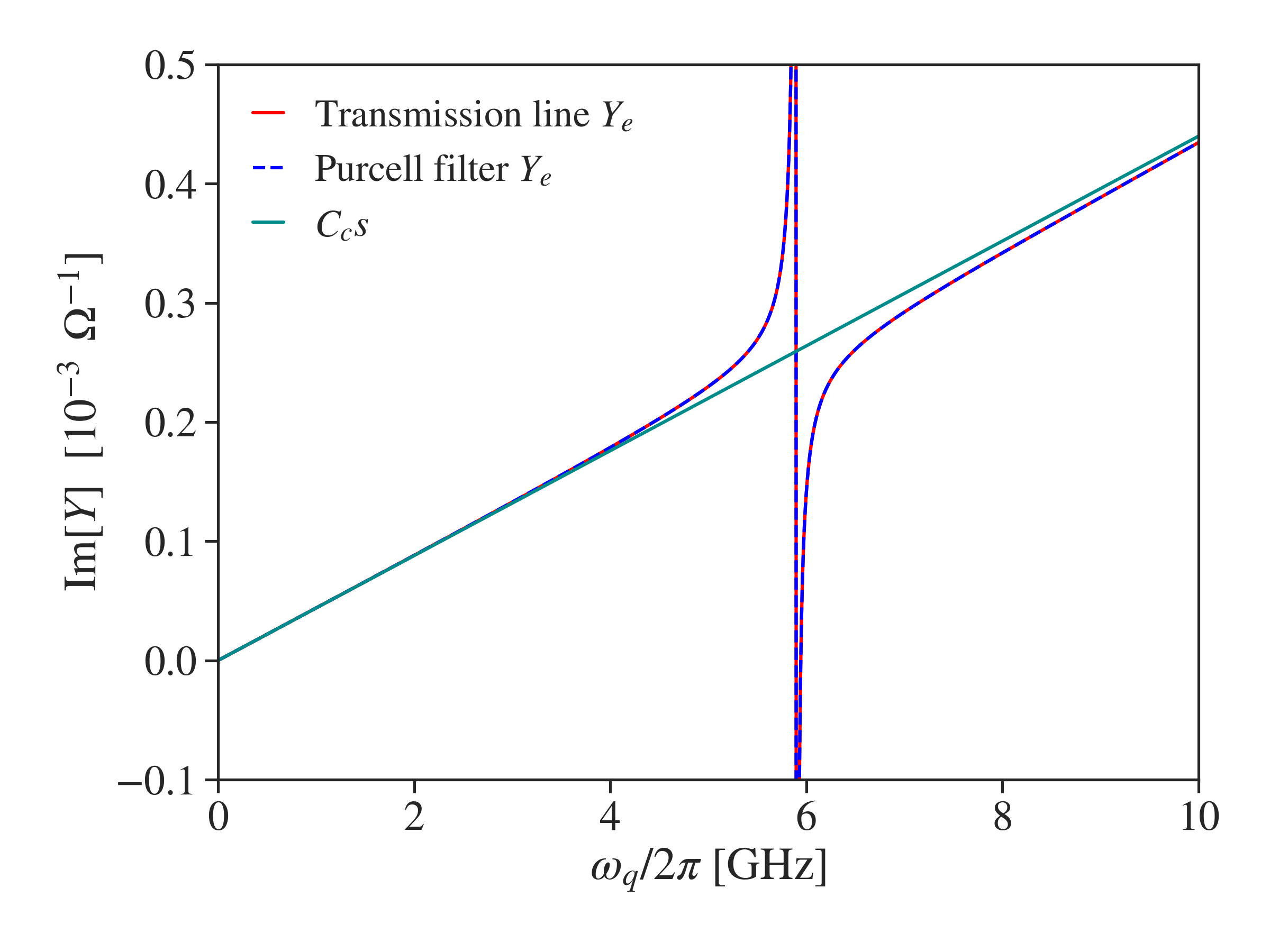}
    \caption{$\Im[Y_e(i\omega_q)]$ for the circuits depicted in Figs. \tlinecirc and \purcellcirc. At low frequency, $\Im[Y_e(i\omega_q)]$ is fit well by  $1/L_e=0$ and $C_e = C_c$. As $\omega_q/2 \pi$ approaches the lowest resonator mode, resonant parts of $Y_e$ become large.}
    \label{fig:c_env_fit}
\end{figure}

In order to apply the single port admittance model for transmon relaxation time \cite{Esteve1986, Martinis2003, Houck2008, Neeley2008} in section \ref{sec:purcell}, we need to find the leading reactances in Eq. \ref{eqn:admittance} for the admittance $Y_e$ seen by the transmon. For frequencies lower than the poles and zeros of $Y_e$, the capacitive and inductive terms dominate the behavior of $\Im[Y_e]$. Fig. \ref{fig:c_env_fit} shows that the model $C_c s$ fits very well to the low frequency behavior of $\Im[Y_e]$ for the circuits depicted in Figs. \tlinecirc and \purcellcirc. We conclude that $C_e\approx C_c$ and $\frac{1}{L_e}\approx 0$.

\section{Convergence}\label{app:convergence}
\begin{figure}
    \centering
    \includegraphics[scale=0.4]{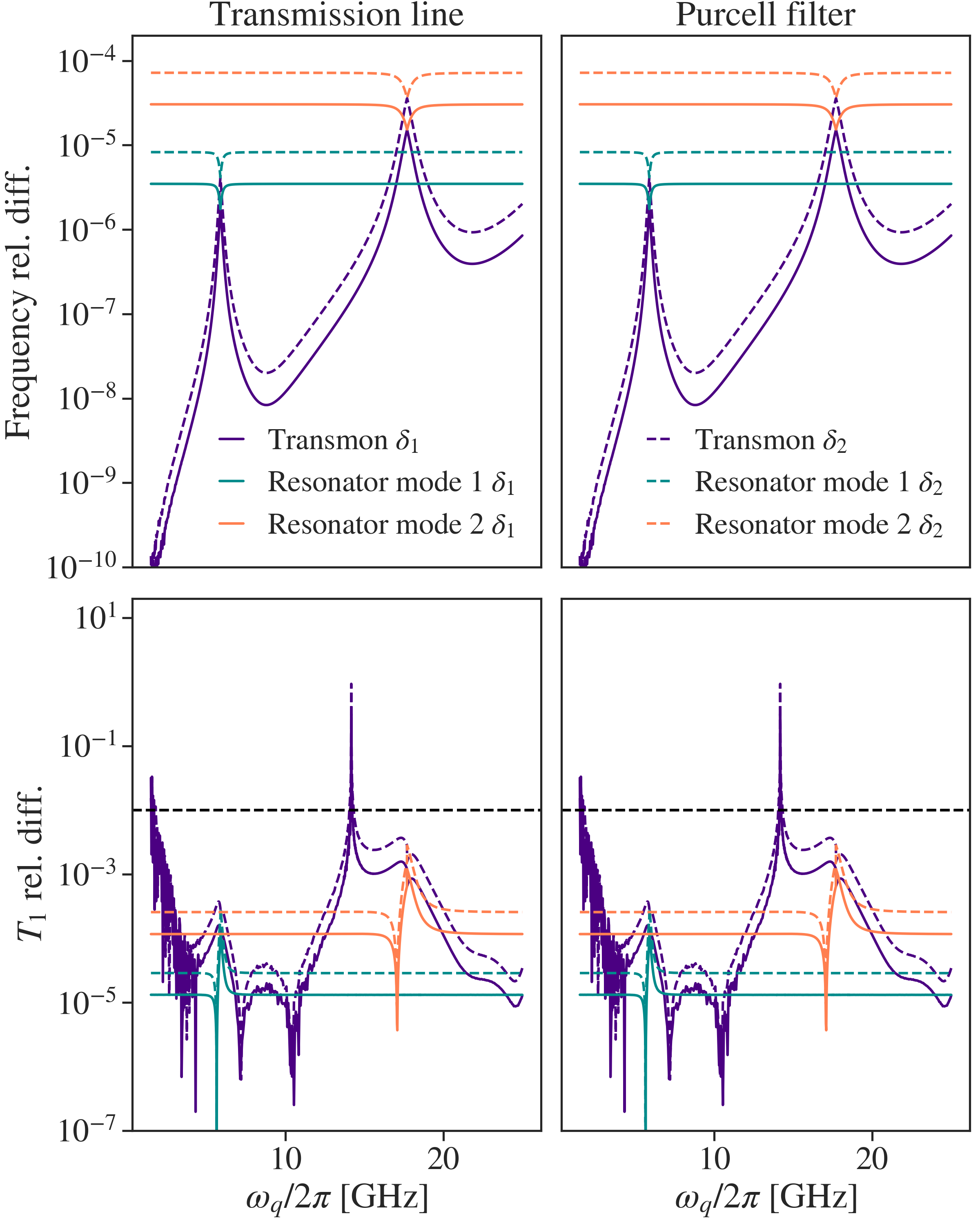}
    \caption{Frequency and relaxation time relative differences to the result for $\delta_0$ for the modes of the circuits depicted in Figs. \tlinecirc and \purcellcirc.}
    \label{fig:tline_pfilter_convergence}
\end{figure}

\begin{figure}
    \centering
    \includegraphics[scale=0.4]{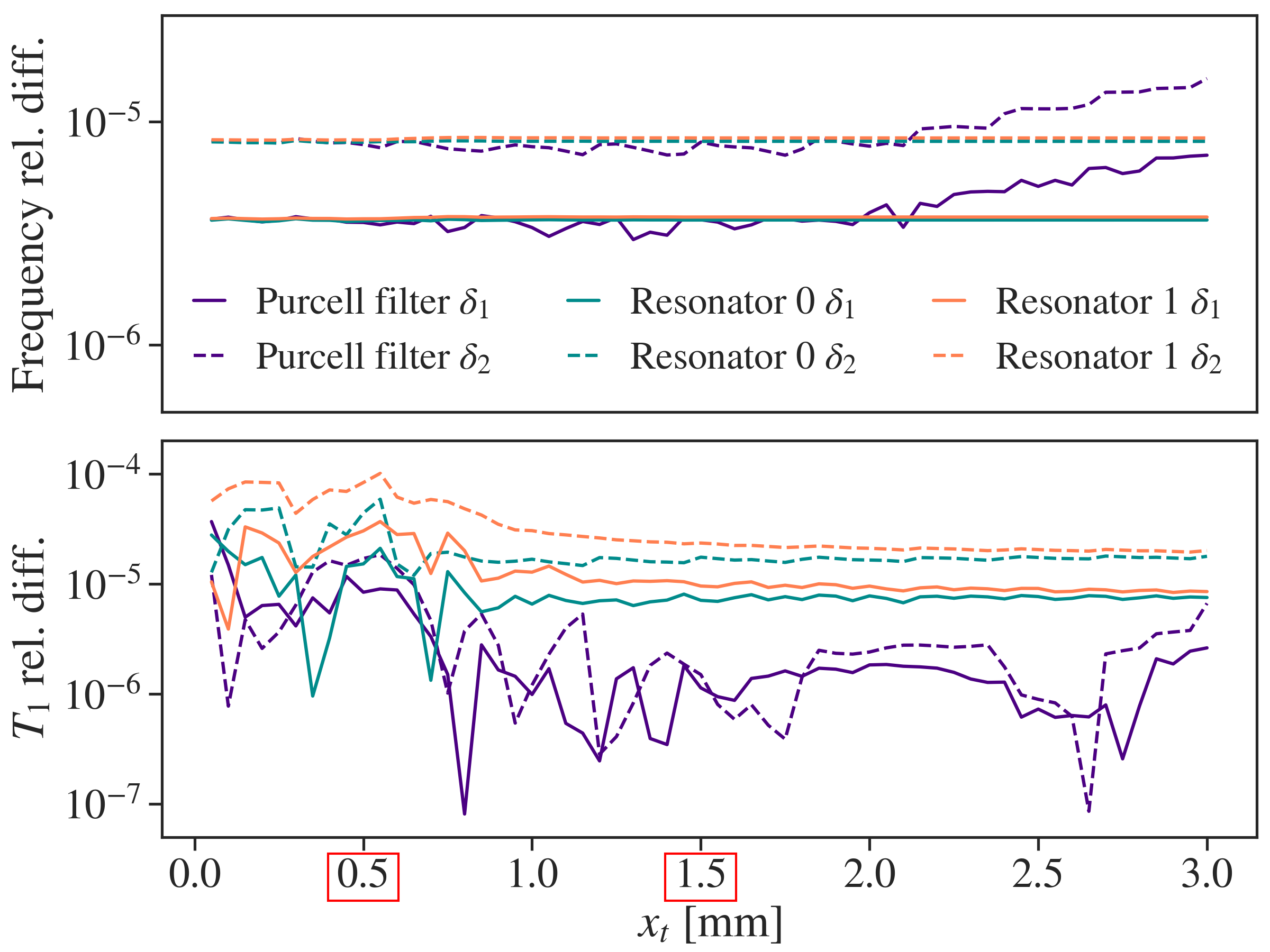}
    \caption{Frequency and relaxation time relative differences to the result for $\delta_0$ for the modes of the circuit depicted in Fig. \tworescirc.}
    \label{fig:hybrdization_convergence}
\end{figure}

In section \ref{sec:modeling}, we approximate transmission lines as $LC$ ladders. In this section, we investigate the error in this approximation as a function of the length scale of the discretization. We choose three discretization lengths $\delta_0 = \SI{40}{\micro\meter}$, $\delta_1 = \SI{50}{\micro\meter}$, $\delta_2 = \SI{60}{\micro\meter}$ and compute the complex frequencies and eigenmodes for each length and for each of the three circuits shown in Fig. \ref{fig:circs}. Fig. \ref{fig:tline_pfilter_convergence} shows the relative differences to the $\delta_0$ result for the frequency and relaxation time for $\delta_1$ and $\delta_2$ as a function of bare qubit frequency for the transmission line and Purcell filter circuits. Fig. \ref{fig:hybrdization_convergence} shows the same quantities for the two resonator circuit as a function of $x_t$.

For all three circuits, the relative frequency difference to the $\delta_0$ result is always less than $0.1\%$ and improves from $\delta_2$ to $\delta_1$. The relative difference of the relaxation times is less than $1\%$ everywhere except where the relaxation time itself is very long. Even in this regime, the errors are small enough that they do not affect our interpretation of Fig. \ref{fig:complex_frequencies}. Furthermore, the relative difference of relaxation times to the $\delta_0$ result improves from $\delta_2$ to $\delta_1$. All results presented in the main body of the paper were computed with a discretization length of $\delta_1 = \SI{50}{\micro\meter}$.

\bibliography{bibliography}
\bibliographystyle{ieeetr}

\end{document}